\documentclass[12pt]{iopart}

\newcommand{\tfrac}{\textstyle\frac}
\newcommand{\der}{\mathrm{d}}
\newcommand{\deq}{\equiv}

\begin{document}
\jl{6}

\title
  {Rotating perfect fluid sources of the NUT metric}

\author{Michael Bradley\dag, Gyula Fodor\ddag, 
  L\'{a}szl\'{o} \'{A}.\ Gergely\ddag\P, Mattias Marklund\dag\  
  and Zolt\'{a}n Perj\'{e}s\ddag}

\address{\dag\ Department of Plasma Physics, Ume{\aa} University,
  S--901 87 Ume{\aa}, Sweden}

\address{\ddag\ KFKI Research Institute for Particle and Nuclear
  Physics, Budapest 114, P.O.Box 49, H--1525 Hungary}

\address{\P\ Laboratoire de Physique Th\'{e}orique, 
  Universit\'{e} Louis Pasteur, 3--5 rue de l'Universit\'{e}, 67084 
  Strasbourg Cedex, France}

 
\begin{abstract} 
Locally rotationally symmetric perfect fluid solutions of
Einstein's gravitational equations are matched along the 
hypersurface of vanishing pressure with the NUT metric. 
These rigidly rotating fluids are interpreted as sources for the
vacuum exterior which consists only of a stationary region of the
Taub-NUT space-time. The solution of the matching conditions 
leaves generally three parameters in the global solution. 
Examples of perfect fluid sources are discussed.
\end{abstract}
 
\pacs{04.20.Cv, 04.20.Jb, 04.40.Dg, 97.60.-s}
 
\section{Introduction}
 
Matching fluids to vacuum exteriors has been a vexed 
question for many years in general relativity. 
Here we present a number
of global solutions to Einstein's field equations representing 
rotating fluids matched to the vacuum NUT metric. The physical 
significance of the latter has also been in the center
of discussion, with at least three differing interpretations 
in the literature \cite{Bonnor,Misner,Demianski-Newman}. 
The intepretation of Demianski and Newman has recently been
revisited by Lynden--Bell and Nouri--Zonoz \cite{Lynden-Bell}.

A recent paper by Mars and Senovilla \cite{MaSe} emphasizes 
the difficulty of matching an axisymmetric rotating interior 
body to a vacuum exterior. They show
that the problem becomes overdetermined when both the
continuity of the induced metric and of the extrinsic curvature of the
junction hypersurface is imposed. Due to the overdetermination it 
is hard to find global
space-times consisting of a rotating core and vacuum exterior.
The weaker requirement of 
continuity of the metric alone was discussed by Shaud and Pfister
\cite{Pfister}. 
 
The present paper was motivated by the structural similarity of 
some rotating perfect fluid solutions 
\cite{Marklund,PFGM,CD,BradleyMarklund}
and the NUT metric \cite{NUT}. We solve the junction conditions
for the radius $r=r_s$ and the parameters of the NUT metric.
 
We cut the perfect fluid space-time  
along the hypersurface of zero pressure, this being an obvious 
requirement for matching with a vacuum solution. In fact, 
the junction conditions single out the zero pressure surface as the 
unique matching surface. 
Then we apply the Darmois--Israel matching procedure 
\cite{Darmois,Israel}. 
The continuity of both the first and second fundamental forms 
impose relations among the parameters of the two solutions.

The fluid space-times are locally rotationally symmetric (LRS), 
i.e., they are invariant under a subgroup of the Lorentz 
group. Further, the LRS class under study has non-zero vorticity
but zero shear and expansion of the fluid. The general system of 
equations determining the properties of the fluid has been 
presented in many forms \cite{Marklund,CD,Stewart-Ellis}.

In section \ref{secLRS}, we briefly review the the theory of LRS 
perfect fluid space-times. We also state the properties of 
the NUT space-time. Next, in section \ref{secjunc} we 
derive the general junction conditions, and finally in section 
\ref{secex} we discuss some examples of the matching procedure 
and the matching space-times. 

Our metric signature convention is $\left( +---\right)$, and
we choose the gravitational constant $G = (8\pi)^{-1}$.

\section{Properties of the LRS perfect fluid and the NUT
  space-times}\label{secLRS}

The locally rotationally symmetric (LRS) space-times are defined to 
be invariant under a spatial isotropy subgroup of the Lorentz group. 
This means that the subgroup is either 1- or 3-dimensional. The 
LRS space-times may be divided into three distinct classes
\cite{Marklund,Stewart-Ellis,vanElst-Ellis}. LRS models were also 
studied by Cahen and Defrise \cite{CD}. In this paper we will 
make use of class I according to the Stewart--Ellis classification
\cite{Stewart-Ellis}. In the generic case, this class is invariant 
under a 1-dimensional isotropy group. 

We choose the following Lorentz tetrad:
\begin{equation} \label{lrstet}
\begin{array}{l}   
  \omega^0 = \displaystyle\frac{\Omega\delta^2}{L}%
             (\der T + 2L\cos\Theta\,\der\Phi)  \\[3mm]
  \omega^1 = \displaystyle\frac{\der R}{y}  \\[3mm]
  \omega^2 = \delta\,\der\Theta             \\
  \omega^3 = \delta\sin\Theta\,\der\Phi \ .   
\end{array}
\end{equation}
where $L$ is a constant and 
\begin{equation}
\delta^{-2} \deq -p+2a\kappa+\kappa^2-\Omega^2 \ . \label{delta}
\end{equation}
Here $p$ is the pressure and, denoting the Ricci rotation 
coefficients by $\gamma_{ijk}$,  
\begin{equation}\label{lrsrot}
  \Omega= - \gamma_{023}\ , \quad \kappa=-\gamma_{122}=-\gamma_{133}
  \ , \quad a=-\gamma_{010}
\end{equation}
are the vorticity, the spatial divergence of the 
basis vector $e_1$ dual to $\omega^1$, and the
acceleration, respectively.\footnote{The Ricci rotation 
  coefficients $\gamma_{ijk}$ satisfy Cartan's first 
  equation $\der\omega^i = 
    \gamma^i\!_{jk}\omega^j\wedge\omega^k$.} 
The function $y$ is determined by the choice of 
the radial coordinate $R$.

The functions in the tetrad depends only on the $R$ coordinate, 
and they are subject to the radial equations 
\begin{equation} \label{LRSeq}
 \begin{array}{l}  
  \displaystyle y\frac{d\Omega}{dR} = (2\kappa - a)\Omega     \\[3mm]
  \displaystyle y\frac{d\kappa}{dR} = \tfrac12(\mu+p) - a\kappa 
     - \Omega^2 + \kappa^2\\[3mm]
  \displaystyle y\frac{da}{dR}      = -\tfrac12(\mu+3p) 
      + a^2 + 2a\kappa +2\Omega^2   \\[3mm]
  \displaystyle  y\frac{dp}{dR}      = a(\mu+p)
 \end{array}                
\end{equation}
where $\mu$ is the energy-density of the fluid.

The static limit is characterized by $\Omega=0$ (note that 
in this case the tetrad (\ref{lrstet}) can not be used). 
This subclass includes all static spherically symmetric 
perfect fluid space-times \cite{Marklund-Bradley},  
e.g., the Tolman \cite{Tolman} and interior Schwarzschild 
solutions.

If we choose our coordinate \label{page:coordchoice}
$R$ such that $\delta^2 = R^2 + L^2$, 
then from (\ref{delta}) and (\ref{LRSeq}) 
$y = -\kappa(R^2+L^2)/R$. Since $p$ is determined algebraically
from equation (\ref{delta}), we may discard the fourth of the 
above radial equations. 

The metric (\ref{lrstet}) bears a striking similarity to that 
of the stationary region of the Taub--NUT
vacuum space-time \cite{Misner,NUT,Taub}, which in a Lorentz 
frame takes the form
\begin{equation}\label{nuttet}
\begin{array}{l}
  \tilde\omega^0 = h(\der t + 2\ell\cos\theta\,\der\varphi) \\[2mm]
  \tilde\omega^1 = \displaystyle\frac{\der r}{h}     \\[3mm]
  \tilde\omega^2 = (r^2 + \ell^2)^{1/2}\,\der\theta        \\[2mm]
  \tilde\omega^3 = (r^2 + \ell^2)^{1/2} \sin\theta\,\der\varphi 
\end{array}
\end{equation}
where
\begin{equation}
  h^2 = \frac{r^2 - 2mr - \ell^2}{r^2 + \ell^2}
\end{equation}
and $m$ and $\ell$ are constants. Using Cartan's first equation, we obtain
the Ricci rotation coefficients, among which
\begin{equation}\label{nutrot}
\begin{array}{l} 
  \tilde\gamma_{010} = \displaystyle%
  \frac{mr^2 - m\ell^2 + 2r\ell^2}%
     {(r^2 + \ell^2)^{3/2}(r^2 - \ell^2-2mr)^{1/2}} \\[4mm]
  \tilde\gamma_{122} = \tilde\gamma_{133} = \displaystyle%
  \frac{r(r^2 - \ell^2 - 2mr)^{1/2}}{(r^2 + \ell^2)^{3/2}} \ . 
\end{array}
\end{equation}
are needed for matching.

\section{The junction conditions between the LRS and NUT
  space-times}\label{secjunc}

The general junction conditions can be stated as follows: 
Find two isometric imbeddings of the matching surface 
$\mathcal{S}$ (with metric $\der s^2_\mathcal{S}$) into the given 
``interior'' and ``exterior'' space-times respectively, 
such that the induced extrinsic curvatures of $\mathcal{S}$ 
may be equated with each other. Thus, the following should 
hold on the matching 3-surface:
\begin{equation}
  K_{ij} = \tilde{K}_{ij} \ , \quad 
  \der s^2|_\mathcal{S} = \der{\tilde s}^2|_\mathcal{S} \ .
\end{equation}
If the surface $\mathcal{S}$ is time-like, choosing a 
normal $n$ to the surface yields the extrinsic curvature 
$K_{ij} = h_i^kh_j^ln_{(k;l)}$ (and analogously for 
$\tilde K$), where $h_{ij} = g_{ij} + n_in_j$ is the 
projector onto $\mathcal{S}$.

We choose the surface $\mathcal{S}$ such that $\omega^1$ 
and $\tilde\omega^1$ are normals to the imbedded surface 
in the two space-time regions. Thus, we take $n = \omega^1$, 
i.e., $n_i = \delta_i^1$. This implies that $\mathcal{S}$
is a surface of constant radius $r$ or $R$.

The non-trivial tetrad components of the extrinsic 
curvature are given as the Ricci rotation coefficients:
\begin{equation}
  K_{\alpha\beta} = \gamma_{1(\alpha\beta)} \ ,
\end{equation}
where $\alpha, \beta = 0, 2, 3$. Then the matching conditions are
\begin{eqnarray}
  \gamma_{1(\alpha\beta)} = {\tilde\gamma}_{1(\alpha\beta)} 
  \label{matching1} \\
  \omega^\alpha|_\mathcal{S} = \tilde\omega^{\alpha}|_\mathcal{S} \ .
  \label{matching2}
\end{eqnarray}
Making the coordinate choice $R^2 = \delta^2 - L^2$ in the LRS case,
and using equations (\ref{lrstet}), (\ref{lrsrot}), (\ref{nuttet}), 
and (\ref{nutrot}), we find
\begin{eqnarray}
  \ell^2 = \frac{\Omega_s^2r_s^2}{\kappa_s^2}  \label{junct1}\\ \bs
  \kappa_s = -\frac{r_s(r_s^2 - \ell^2 - 2mr_s)^{1/2}}%
                   {(r_s^2 + \ell^2)^{3/2}}%
    \label{junct2}\\ \bs
  a_s = -\frac{mr_s^2 - m\ell^2 + 2r_s\ell^2}%
              {(r_s^2 + \ell^2)^{3/2}(r_s^2 - \ell^2 - 2mr_s)^{1/2}} 
    \ , \label{junct3}
\end{eqnarray}
with $R_s = r_s$ and $L = \ell$, and where the subscript $s$ 
is used to denote quantities on $\mathcal{S}$.
We choose $t = T$, $\theta = \Theta$, and $\varphi = \Phi$ 
on the junction surface $\mathcal{S}$. 
Using the above given radial coordinate definition, equation 
(\ref{delta}) is found to be trivially satisfied on the 
zero-pressure surface due to the junction conditions. 
Therefore it may be used in place of, say, equation 
(\ref{junct3}). We solve equations (\ref{junct1}), 
(\ref{junct2}) and (\ref{delta})
for the NUT parameters and the matching radius, and obtain
\begin{eqnarray}
  \ell^2    = \frac{\Omega_s^2}{(\kappa_s^2+\Omega_s^2)%
        (2a_s\kappa_s + \kappa_s^2 - \Omega_s^2)} 
        \label{junct1'} \\ \bs 
  mr_s = \frac{\kappa_s(a_s\kappa_s^2 - a_s\Omega_s^2 
           - 2\kappa_s\Omega_s^2)}{(\kappa_s^2 + \Omega_s^2)%
        (2a_s\kappa_s + \kappa_s^2 - \Omega_s^2)^2} 
        \label{junct2'} \\ \ms
  r_s^2  = \frac{\kappa_s^2}{(\kappa_s^2 + \Omega_s^2)%
        (2a_s\kappa_s + \kappa_s^2 - \Omega_s^2)} \label{junct3'}
\end{eqnarray}
with the subsidiary condition $2a_s\kappa_s + \kappa_s^2 
  - \Omega_s^2 > 0$ (and $mr_s > 0$, if we want a positive 
mass parameter). In order to confine the metric to the 
stationary domain of the Taub--NUT space-time, then 
the matching radius $r_s$ should be larger than the limiting
radius of the Taub region
$r_{\rm T} = m + \sqrt{m^2 + \ell^2}$, a condition expressed 
as $r_s^2 - \ell^2 - 2mr_s > 0$. This is satisfied, since
\begin{equation}
  r_s^2 - \ell^2 - 2mr_s = \frac{\kappa_s^2 
     + \Omega_s^2}{(2a_s\kappa_s 
     + \kappa_s^2 - \Omega_s^2 )^2} \ .
\end{equation}

The static limit ($\Omega = 0$) of the fluid requires $\ell = 0$. 
One hence gets a matching to the exterior Schwarzschild solution 
in this limit. 

In the generic case, having chosen an equation of 
state $\mu=\mu(p)$, the integral of equations 
(\ref{LRSeq}) contains three constants after the 
coordinate $R$ has been chosen. Therefore there remain 
three independent parameters in the matching conditions.
Finding the solution in the generic case remains difficult 
even with the equation of state specified. 
Simplifying assumptions yield metrics with a reduced number
of parameters. This will imply that the NUT parameters and 
the junction radius are not independent parameters. 
In certain cases, with examples given in the 
next section, it will not be possible to satisfy 
all the positivity conditions. 

\section{Explicit global solutions}\label{secex}

When joining the two metrics, the radial coordinate of 
the fluid has to be specified. The coordinate choice 
$R^2 = \delta^2 - L^2$ is convenient for our matching 
procedure, but it covers only the region outside the surface
$\delta = L$. Thus, in the fluid interior, it is more 
appropriate to use the function $\delta$ as the radial 
coordinate. In each case presented below, the metric may be 
reconstructed using the tetrads (\ref{lrstet}) and 
(\ref{nuttet}).

\subsection{Incompressible fluid with constant vorticity} 

Assuming that the vorticity $\Omega \deq \Omega_s$ and 
the density $\mu \deq \mu_s$ are constants, equations 
(\ref{LRSeq}) yield $a = 2\kappa$. Furthermore, the density, 
pressure and acceleration are given in the form 
\cite{Marklund,PFGM}
\begin{equation}
  \mu_s = 6\Omega_s^2 \ , \quad 
  p = \frac{4}{\delta^2} - 6\Omega_s^2 \ ,
  \quad a^2 = \frac{4}{\delta^2} - {4}\Omega_s^2 \ .
\end{equation}

The coordinate choice $\delta^2 = R^2 + L^2$  gives the junction
conditions (\ref{junct1'})--(\ref{junct3'}) in the simple form
\begin{equation}
  \ell^2 = \frac{4}{9\Omega_s^2} \ , \quad mr_s = -\frac{4}{9\Omega_s^2}
  \ , \quad r_s^2 = \frac{2}{9\Omega_s^2} \ .
\end{equation}
There is only one free parameter, $\Omega_s$. 
We see further that the mass is negative. It is straightforward
to show that this defect can 
not be removed by including a cosmological constant.

\subsection{A fluid with variable density and constant vorticity}

As in the previous case, this fluid has $\Omega \deq \Omega_s$ a
constant and $a=2\kappa$, but with an equation of state
\begin{equation}
  p = 12\Omega_s^2 - 3\mu 
    + A(\mu - 6\Omega_s^2)^{3/2} \ ,
\end{equation}
with $\mu > 6\Omega_s^2$ guaranteeing the positivity of the 
speed of sound. Here $A$ is a constant.
The density, pressure and the acceleration may be expressed 
as \cite{Marklund}
\begin{eqnarray}
  \mu = \left(\frac{4}{A\delta^2}\right)^{2/3} + 6\Omega_s^2 \\ \bs
  p   = \frac{4}{\delta^2} - 
        3\left(\frac{4}{A\delta^2}\right)^{2/3} - 6\Omega_s^2 \\ \bs
  a^2 = \frac{4}{\delta^2} - 
        \frac{12}{5}\left(\frac{4}{A\delta^2}\right)^{2/3}
        - 4\Omega_s^2 \ .
\end{eqnarray}
On the junction surface ${\mathcal{S}}$, with the coordinate 
choices as before, we have the matching 
conditions
\begin{eqnarray}
  \ell^2     = \frac{5\Omega_s^2\delta_s^4}{1 + 6\Omega_s^2\delta_s^2} 
               \\ \bs  
  mr_s       = \frac25\frac{\delta_s^2(1 + \Omega_s^2\delta_s^2)%
                            (1 - 19\Omega_s^2\delta_s^2)}
                 {1 + 6\Omega_s^2\delta_s^2} \\ \bs
  r_s^2      = \delta_s^2\frac{1 
               + \Omega_s^2\delta_s^2}{1 + 6\Omega_s^2\delta_s^2}
\end{eqnarray}
where  
\begin{eqnarray}
  f = 18\left(\sqrt{\tfrac23 + (A\Omega_s)^2} 
         - A\Omega_s\right) \\ \bs
  \delta_s^2 = \frac{f^{1/3}}{3A\Omega_s^2} - 
               \frac{2}{A\Omega_s^3f^{1/3}} 
              + \frac{2}{3\Omega_s^2} \ . 
\end{eqnarray}
The positivity of the mass $m$ is guaranteed by the inequality
\begin{equation}
  \vert A\Omega_s\vert < 
  \left(\frac{2}{19}\right)^{1/2}
  \left(\frac{57}{35}\right)^{3/2} \approx 0.67 \ .
\end{equation}

\subsection{Purely electric Weyl tensor}
  
The condition that the Weyl tensor is purely electric 
\cite{Marklund,BradleyMarklund} yields that $a=\kappa$ and 
\begin{eqnarray} 
  p        = -\frac{1}{2}\mu_s 
             + \frac{1}{8D - 2}\delta^{-2} \\ \bs
  \Omega^2 = \frac{1 - 2D}{8D - 2}{\delta^{-2}}     \\ \bs
  a^2      = -\frac{1}{6}\mu_s 
             + \frac{2D}{8D - 2}\delta^{-2} \ ,
\end{eqnarray}
where $\mu_s$ is a positive constant. The equation of state is given by
\begin{equation}
  \mu=p+\mu_s\ .
\end{equation}
The reality of these quantities implies $1/4<D<1/2$. 

Using the junction conditions (\ref{junct1'})--(\ref{junct3'}), we
find the following expressions for the NUT parameters and the junction 
radius
\begin{eqnarray}
  \ell^2 = \frac{3(1 - 2D)}{8\mu_s(1 - 4D)(1 - 3D)}  \\ \bs
  mr_s   = \frac{(468D^2 - 288D + 43)(12D - 5)}%
              {24\mu_s(11 - 30D)(1 - 4D)^2(1 - 3D)}  \\ \bs
  r_s^2  = \frac{30D - 11}{8\mu_s(1 - 4D)(1 - 3D)} \ .
\end{eqnarray}
The positivity of the mass parameter forces the value of $D$ 
to lie in the open interval $(5/12,1/2)$. Note that there 
are two free parameters, $D$ and $\mu_s$, in the above 
junction relations.

\section{Discussion}

In the previous section we have obtained global solutions to 
Einstein's equations, with the interiors representing a 
rigidly rotating perfect fluid each with a given equation 
of state. The matching with the exterior NUT space-time is 
along a freely chosen surface, selected by the appropriate 
choice of the parameters such that the pressure vanishes 
on this surface.

Because of the structure of the metric, the fluid interior 
shares the counter-intuitive properties of the exterior NUT 
space-time, e.g., closed time-like curves. In general, 
the LRS class I fluids all have curvature singularities 
at the center $\delta = 0$ (there may exist solutions which 
can not be extended to $\delta = 0$, although their physical 
status is quite unclear). 

We emphasize that these fluids do not have a unique axis of 
rotation, instead they appear to rotate about every 
radial direction. 

\ack
 
This research has been supported by OTKA grants T17176, 
D23744 and T022563. We also acknowledge the support of 
the Royal Swedish Academy of Sciences, the Hungarian 
Academy of Sciences, the Swedish Natural Science Research 
Council and the Hungarian State E\"{o}tv\"{o}s Fellowship.

\end{document}